\documentclass[prb,reprint,superscriptaddress,floatfix,aps,onecolumn]{revtex4-2}
\usepackage{amsmath}
\usepackage{amsfonts}
\usepackage{graphicx}
 \usepackage{physics}
 \usepackage[usenames]{color}

 \def\be{\begin{equation}} \def\ee{\end{equation}}
\def\bea{\begin{eqnarray}} \def\eea{\end{eqnarray}}

\begin{document}

\title{Quasi-Discrete Time Crystals in the quasiperiodically driven Lipkin-Meshkov-Glick model}

\author{Sk Anisur}
\affiliation{Harish-Chandra Research Institute, a CI of Homi Bhabha National Institute, Chhatnag Road, Jhunsi, Allahabad 211019}
	
\author{W. Vincent Liu}
\email{wvliu@pitt.edu}
\affiliation{Department of Physics and Astronomy, University of Pittsburgh, Pittsburgh, PA 15260, USA}

\author{Sayan Choudhury}
\email{sayanchoudhury@hri.res.in}
\affiliation{Harish-Chandra Research Institute, a CI of Homi Bhabha National Institute, Chhatnag Road, Jhunsi, Allahabad 211019}
\affiliation{Department of Physics and Astronomy, University of Pittsburgh, Pittsburgh, PA 15260, USA}

\date{\today}
	
\begin{abstract}
A discrete time crystal (DTC) is a remarkable non-equilibrium phase of matter characterized by the persistent sub-harmonic oscillations of physical observables in periodically driven many-body systems. Motivated by the question of whether such a temporal periodic order can persist when the drive becomes aperiodic, we investigate the dynamics of a Lipkin-Meshkov-Glick model under quasiperiodic Thue-Morse (TM) driving. Intriguingly, this infinite-range-interacting spin system can host  ``quasi- discrete time crystal" (quasi-DTC) phases characterized by periodic oscillations of the magnetization. We demonstrate that our model can host the quasi-DTC analog of both period-doubling DTCs as well as higher-order DTCs. These quasi-DTCs are robust to various perturbations, and they originate from the interplay of ``all-to-all" interactions and the recursive structure of the TM sequence. Our results suggest that quasi-periodic driving protocols can provide a promising route for realizing novel non-equilibrium phases of matter in long-range interacting systems.
	  
\end{abstract}
	
\maketitle
	
\section{Introduction} 

In recent years, periodic driving has emerged as a powerful tool for the coherent control of quantum many-body systems \cite{eckardt2005superfluid,holthaus2015floquet,bukov2015universal,eckardt2017colloquium,oka2019floquet,moessner2017equilibration,rudner2020band}. Periodically driven (Floquet) systems provide a versatile platform for designing quantum many-body states which may not have any equilibrium analog. A particularly intriguing example of such a non-equilibrium phase of matter is a discrete time crystal (DTC) \cite{sacha2018review,sacha2020book,khemani2019review,nayak2019review}. A DTC exhibits spontaneous discrete time-translation-symmetry-breaking (TTSB), and a consequent stable sub-harmonic response of physical observables. Following the first pioneering theoretical proposals \cite{sachapra2015,sondhi2016prl,sondhi2016prb,nayak2016prl,yao2017prl}, DTCs have now been realized in several experimental platforms \cite{monroe2017nature,lukin2017nature,barrett2018prl,mahesh2018prl,apal2019arxiv,kessler2020continuous,randall2021observation,kessler2020observation,mi2021observation}. \\

While Floquet engineering protocols are remarkably powerful, they have an inherent drawback - periodic driving typically heats up many-body systems to a featureless infinite temperature state \cite{d2014long,lazarides2014equilibrium}. To circumvent this
problem, the first realization of a DTC relied crucially on the presence of many-body localization (MBL)~\cite{monroe2017nature}. Many-body localized systems are characterized by an extensive number of emergent local integrals of motion and they do not absorb heat from the drive, thereby providing a robust route to avoid heat death~\cite{sondhi2016prb}. Unfortunately, the requirement of MBL places very restrictive constraints on the strength and range of interactions \cite{gopalakrishnan2020dynamics,abanin2019colloquium,morningstar2021avalanches,kiefer2021slow} This issue has necessitated the search for other pathways to stabilize a DTC~\cite{liu2018prl,fazio2017prb,mizuta2018spatial,yu2019discrete,fazio2019prb,pizzi2019prl,kshetrimayum2020stark,lyu2020eternal,choudhury2021route,pizzi2020time,pizzi2021higher}. Interestingly, some recent studies have demonstrated that a large class of DTCs can be realized in clean long-range interacting systems~\cite{pizzi2021higher,giachetti2023fractal,liu2024higher}. These time crystals are robust to various perturbations that preserve the perfect periodicity of the external drive. However, random temporal disorder leads to a rapid destruction of the DTC order in isolated quantum systems~\cite{cosme2023bridging}. This naturally raises an important question: can a long-lived DTC-like order emerge in an aperiodically driven system?\\

In this work, we answer this question affirmatively by demonstrating the existence of a zoo of DTC-like phases, characterized by periodic temporal order, in the quasi-periodically driven Lipkin-Meshkov-Glick (LMG) model~\cite{lipkin1965validity}. We note that the LMG model is characterized by `all-to-all' interactions and it can exhibit robust DTC order under an appropriate Floquet protocol \cite{pizzi2021higher}. However, quasiperiodically driven systems are generally expected to heat up much faster than their Floquet counterparts, due to the presence of multiple incommensurate frequencies in the drive~\cite{pilatowsky2023complete}. Intriguingly, we find that this chain exhibits persistent oscillations for a wide parameter regime. Furthermore, we establish the stability of this phase in the thermodynamic limit, thus allowing us to avoid any erroneous conclusions that can arise from finite-size numerics~\cite{pizzi2020time}. Finally, we note that this model can be realized in a variety of quantum emulator platforms, thereby making our proposal amenable to near-term experimental realizations.\\

This manuscript is organized as follows. In sec.~\ref{sec:Model}, we describe the model and establish that this system hosts period-doubling and higher-order DTCs in the presence of a periodic drive. Next, we explore the dynamics of this model in the presence of Thue-Morse driving and determine the conditions for realizing period-doubling and higher order quasi-DTCs in sec.~\ref{sec:TMDynamics}. We discuss two potential experimental realizations of our model in sec~\ref{sec:Experiment}. Finally, we conclude in sec.~\ref{sec:Summary} with a summary of our results and a brief outlook for future investigations on quasi-periodically driven many-body systems.

\section{Model and DTC dynamics} 
\label{sec:Model}

\begin{figure}[t]
		\includegraphics[scale=0.28]{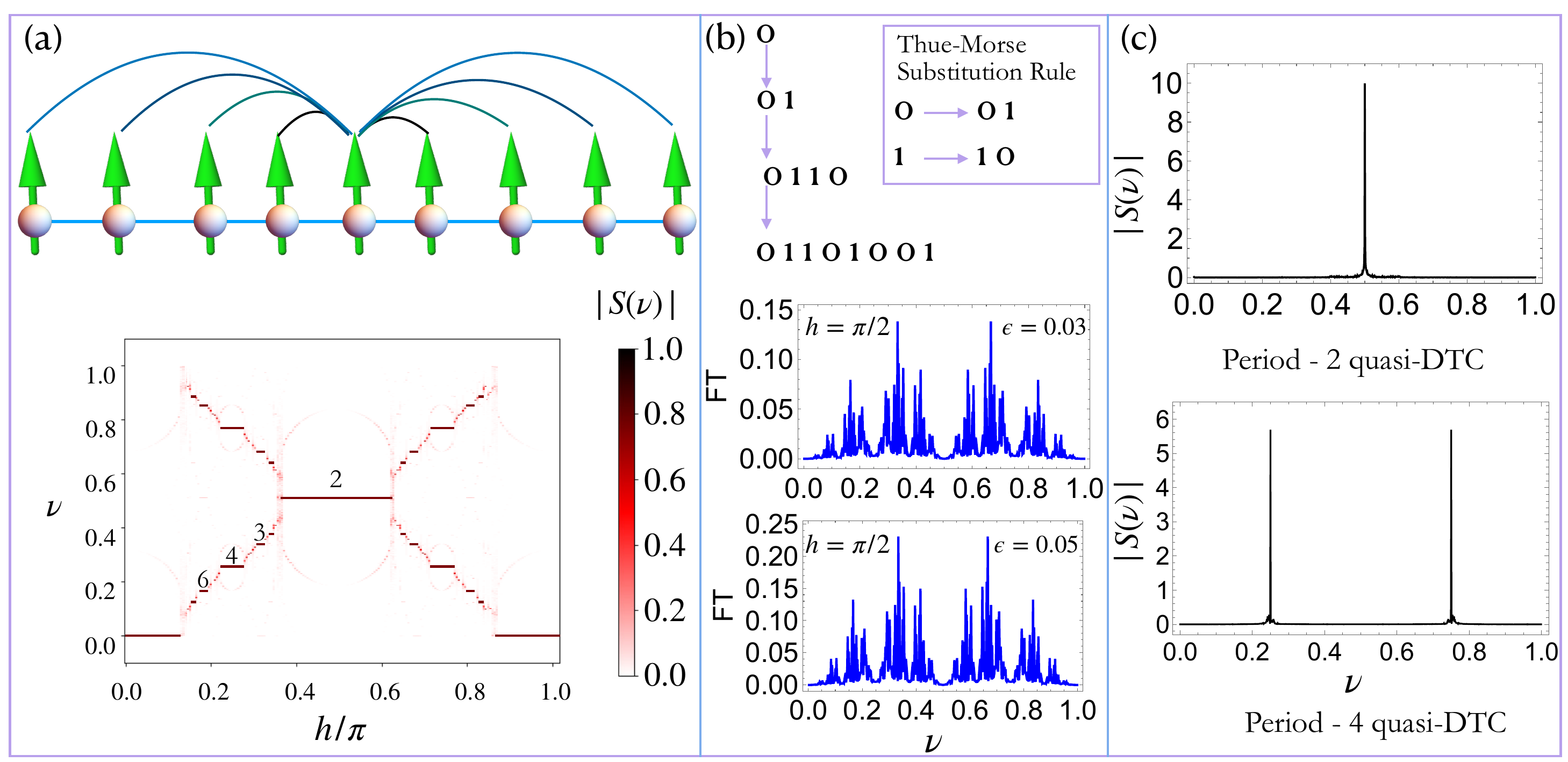}
		\caption{{\bf Time crystals and the Thue-Morse drive sequence:} (a) The top panel shows the schematic of the driven Lipkin-Meshkov-Glick model in Eq.~(\ref{eq:model}). This system is characterized by infinite-range Ising interactions of strength $J$ and is subjected to periodic kicks with a transverse magnetic field strength $\pi h$. The bottom panel shows the density plot of the Fourier transform, $|S(\nu)|$ of the z-magnetization, $S^z(t)$ defined in Eq.~(\ref{eq:order-parameter}) as a function of $h$, when $J=1$. (b) The top panel illustrates the procedure to generate the Thue-Morse (TM) sequence described in Sec.~\ref{sec:TMDynamics}. The bottom panel shows the Fourier transform of the TM kicking term: $\sum_{n}\delta(t-n)\frac{\pi}{m}(\epsilon\tau_n)$, where $\tau_n$ is the $n-$th element of the TM sequence. The plots show the Fourier spectrum for $m=2$ and $\epsilon = 0.03$ and $\epsilon=0.05$. The spectrum for other values of $m$ and $\epsilon$ would be qualitatively similar. (c) The Fourier transform $|S(\nu)|$ of the z-magnetization, corresponding to persistent periodic oscillations with period-2 (top) and period-4 (bottom) in the TM driven LMG chain (Eq.~(\ref{eq:TMmodel})). This phase is dubbed a `quasi-DTC'.}
		\label{fig1}
\end{figure}

We study a driven LMG chain describing the dynamics of $N$ spin-$1/2$ particles mutually interacting through an infinite-range Ising interaction subjected to a time-dependent transverse magnetic field:
\begin{equation}
H = \frac{J}{2N} \sum_{i,j=1}^N \sigma_i^z \sigma_j^z + \sum_{n \in \mathbb{Z}} \delta (t-n) \left[h(1 - \epsilon \tau_n) \sum_i \sigma_i^x \right],
\label{eq:model}
\end{equation} 
where $\sigma_i^{\gamma}$ are the standard Pauli matrices, $\pi h$ is the strength of an external transverse magnetic field, $n \in \mathbb{Z}$, and $\tau_n$ is $n$-th element of a periodic or aperiodic sequence. A schematic of this model is shown in Fig.~\ref{fig1}(a). In the rest of this work, we primarily analyze the time-evolution of the system from the fully $z$-polarized initial state, $\vert \psi (t=0) \rangle = \vert \uparrow \uparrow \uparrow \ldots \uparrow \uparrow \uparrow \rangle$, and characterize the dynamical phases of this system by examining the dynamics of the average $z-$magnetization:
\begin{equation}
S^z(t) = \frac{1}{N} \sum_j \langle \psi \vert \sigma_j^z (t) \vert \psi \rangle, 
\label{eq:order-parameter}
\end{equation} 
We note that the Fourier spectrum of $S^z$ has been conventionally used to identify the DTC~\cite{sondhi2016prl,sondhi2016prb} and time quasicrystal~\cite{dumitrescu2018logarithmically,zhao2019floquet} phases.\\

This model can be directly analyzed in the thermodynamic limit, $N\rightarrow \infty$ by semiclassical methods. To do this, we first re-write the collective spin operators, $S^{\gamma} = \sum_{i=1}^N \sigma_i^{\gamma}$ in terms of standard bosonic operators, $a_{\uparrow}, a_{\downarrow}, a_{\uparrow}^{\dagger}$, and $a_{\downarrow}^{\dagger}$:
\begin{eqnarray}
   S^{x} &=& a_{\uparrow}^{\dagger} a_{\downarrow} + a_{\downarrow}^{\dagger} a_{\uparrow}, \nonumber \\
    S^{y} &=& - i (a_{\uparrow}^{\dagger} a_{\downarrow} - a_{\downarrow}^{\dagger} a_{\uparrow}), \nonumber \\
    S^{z} &=& a_{\uparrow}^{\dagger} a_{\uparrow} - a_{\downarrow}^{\dagger} a_{\downarrow}.
\end{eqnarray}
In the following analysis, we employ the standard definition of the bosonic number operators, $n_{\uparrow} = a_{\uparrow}^{\dagger} a_{\uparrow}$ and $n_{\downarrow} = a_{\downarrow}^{\dagger} a_{\downarrow}$ and obtain the equations of motion of the bosonic operators, $a_{\uparrow}$ and $a_{\downarrow}$:
\begin{eqnarray}
\frac{d a_{\uparrow}}{dt} &=& i [H, a_{\uparrow}]= - h \tau_n \delta(t-n) a _{\downarrow} - \frac{2J}{N} a_{\uparrow} n_{\uparrow},\\
\frac{d a_{\downarrow}}{dt} &=& i [H, a_{\downarrow}]= - h \tau_n \delta(t-n) a _{\uparrow} - \frac{2J}{N} a_{\downarrow} n_{\downarrow}.
\end{eqnarray}
In the $N\rightarrow \infty$ limit, we can substitute $a_{\uparrow} \rightarrow \sqrt{N} \psi_{\uparrow}$ and $a_{\downarrow} \rightarrow \sqrt{N} \psi_{\downarrow}$, where $\psi_{\uparrow}$ and $\psi_{\downarrow}$ are complex fields; this process leads to a Gross-Pitaveskii equation (GPE):
\begin{eqnarray}
\frac{d \psi_{\uparrow}}{dt} &=& -\pi h \tau_n \delta(t-n) \psi _{\downarrow} - 2J |\psi_{\uparrow}|^2 \psi_{\uparrow},\\
\frac{d \psi_{\downarrow}}{dt} &=& -\pi h \tau_n \delta(t-n) \psi _{\uparrow} - 2J |\psi_{\downarrow}|^2 \psi_{\downarrow}.
\end{eqnarray}
We numerically integrate this GPE to study the dynamics of the system. The z-magnetization can then be computed by using:  
\begin{equation}
S^z(t) = \vert \psi_{\uparrow}(t)|^2-\vert \psi_{\downarrow}(t)|^2.
\label{eq:orderparameter}
\end{equation}
Furthermore, we note that $S^x(t)$ and $S^y(t)$ can be obtained using:
\begin{equation}
    S^x = \frac{(\psi_{\uparrow}^*\psi_{\downarrow}+ \psi_{\downarrow}^*\psi_{\uparrow})}{2} \,\,\,\, {\rm and} \,\,\,\,  S^y =\frac{(\psi_{\uparrow}^*\psi_{\downarrow}- \psi_{\downarrow}^*\psi_{\uparrow})}{2i}.
\end{equation}
This leads to the following equations for the spin components:

\begin{eqnarray}
S^x_{n+1} &=&  {\rm Re}\bigg[\left(-S^y_n \cos(\theta_n) + S^z_n \sin(\theta_n) + i S^x_n \right)\exp\left[-i 2J (S^z_n\cos(\theta_n)  + S^y_n \sin(\theta_n))\right]\bigg]\\
S^y_{n+1} &=& {\rm Im}\bigg[\left(-S^y_n \cos(\theta_n) + S^z_n \sin(\theta_n) + i S^x_n\right)\exp\left[-i 2J (S^z_n\cos(\theta_n)  + S^y_n \sin(\theta_n))\right]\bigg],\\
S^z_{n+1} &=& S^y_n \sin(\theta_n) + S^z_n \cos(\theta_n),
\end{eqnarray}

where $\theta_n = h(1 - \epsilon \tau_n)$ and $S^{\alpha}_n = S^{\alpha} (nT)$. We note that these equations can also be obtained by taking the $S \rightarrow \infty$ limit of the Heisenberg equations of motion for the collective spin operators, $S^{\alpha}$~\cite{haake1987classical}.\\

Before analyzing the dynamics of this system under aperiodic driving, we review the properties of this model when the system is subjected to a perfect periodic drive ($\tau_n = 0$). As shown in Fig.~\ref{fig1}(a), this system can host a large zoo of stable DTC phases around $h\sim \pi/m$ for certain values of $m$. These different DTCs correspond to robust TTSB with different sub-harmonic responses. Firstly, we note that this system hosts a robust period-doubling time crystal when $m=2$ and $h_c^{(1)} \le h \le h_c^{(2)}$, where $h_c^{(1)} \sim 0.4$ and $h_c^{(2)} \sim 0.6$ for $J=1$. This DTC originates from the underlying $\mathbb{Z}_2$ symmetry of the model and it has also been observed in many-body localized systems~\cite{sondhi2016prb,sondhi2016prl,yao2017prl}.\\

Intriguingly, this system also hosts a zoo of other DTCs corresponding to persistent periodic oscillations with a period $mT$ where $m>2$. These DTCs have been dubbed Higher order (HO)-DTCs and they arise when $h\sim \pi/m$, for certain values of $m$. These HO-DTCs are characterized by peristent oscillations of physical observables, such as the $z-$magnetization with a period $mT$. Consequently, the Fourier transform of the stroboscopic $z-$magnetization, $S(\nu) = \sum_n S^z(n) \exp(i 2 \pi \nu n)$ exhibits sharp peaks at $\omega = 1/m$ and $\omega = 1-1/m$. Our results are shown in Fig.~\ref{fig1}(a). We focus on $m=3,4,$ and $6$ cases in this work, since these are associated with robust DTC phases. We note that these HO-DTCs are rare and have only been observed in long-range interacting systems~\cite{pizzi2021higher,giachetti2023fractal,liu2024higher} and central spin models~\cite{biswas2025floquet}. Intriguingly, HO-DTCs can be a powerful resource for quantum-enhanced multi-parameter sensing~\cite{biswas2025floquet}. We now proceed to examine the fate of these DTCs in the presence of a quasiperiodic Thue-Morse (TM) drive.\\

\begin{figure*}
		\includegraphics[scale=0.2]{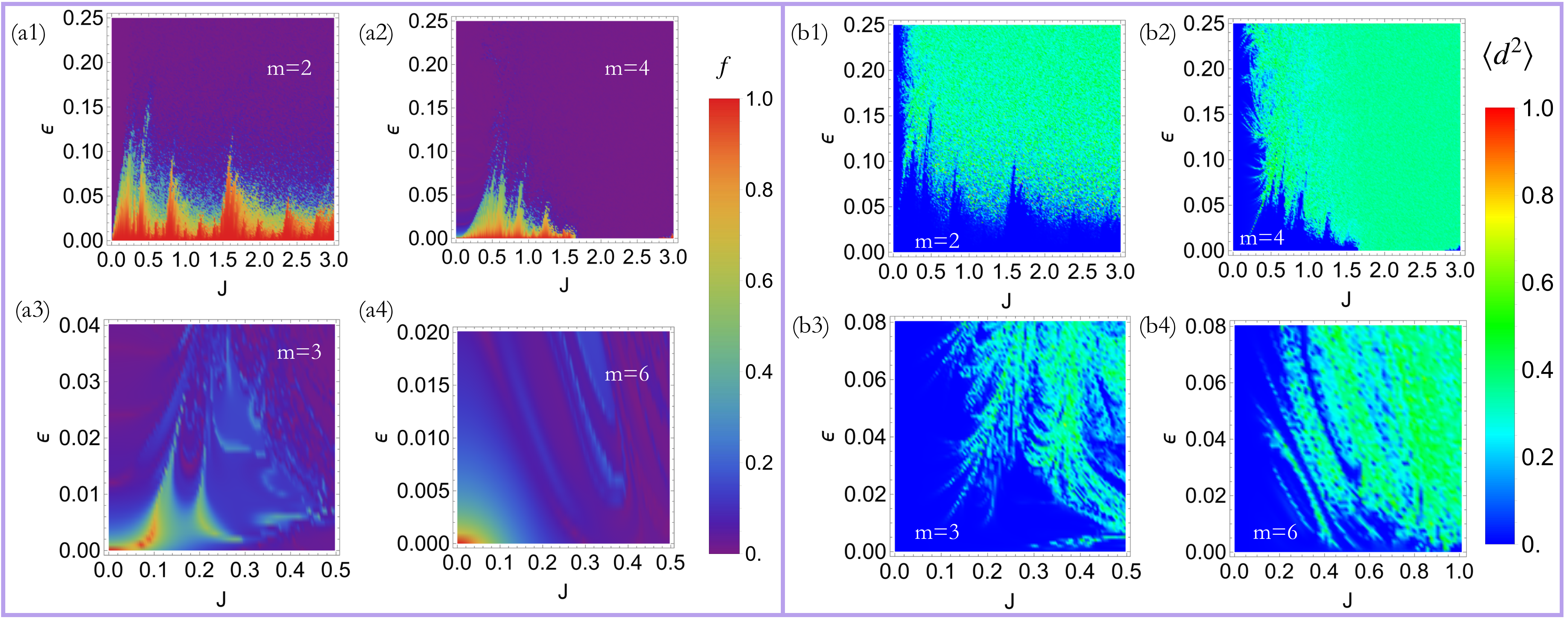}
		\caption{{\bf Dynamical phases of the Thue-Morse driven LMG model} (a1)-(a4): The quasi-DTC fraction (defined in Eq.~(\ref{eq:DTCFrac})) for the Thue-Morse driven LMG model (Eq.~(\ref{eq:TMmodel})) after 500 periods. Robust quasi-DTCs are observed for a wide parameter regime when $m=2$ and $4$, corresponding to period-2 and period-4 quasi-DTCs respectively. Quasi-DTCs are also observed when $m=3$ and $6$, though these are more fragile. (b1)-(b4) The time-averaged decorrelator, $\langle d^2 \rangle$ (defined in Eq.~(\ref{eq:decorr})) for the Thue-Morse driven LMG model after 500 periods. A high (low) value of the decorrelator with $\langle d^2 \rangle \sim 1$ ( $\langle d^2 \rangle \sim 0$) indicates a chaotic (non-chaotic) phase. A comparison with (a1)-(a4) clearly demonstrates that the quasi-DTC phase is associated with a low value of $\langle d^2 \rangle$. The parameters $\epsilon$ and $J$ are defined in Eq.~(\ref{eq:DTCFrac}).}
		\label{fig2}
\end{figure*}

\section{Thue-Morse driving and Quasi-DTC Dynamics} 
\label{sec:TMDynamics}
Quasi-periodic driving protocols provide a promising route to realize non-equilibrium phases of matter beyond the Floquet paradigm~\cite{dumitrescu2018logarithmically,zhao2019floquet,he2025experimental,dutta2025prethermalization}. A celebrated driving protocol in this context is TM driving~\cite{nandy2017aperiodically,zhao2021random,mori2021rigorous,tiwari2024dynamical,tiwari2024periodically,mukherjee2020restoring}, where novel long-lived prethermal phases of matter such as time rondeau crystals can be realized~\cite{zhao2023temporal,moon2024experimental,ma2025stable,kumar2024prethermalization}. Notably, any many-body system would eventually thermalize in the presence of a TM drive, even though this thermalization process may be extremely slow~\cite{pilatowsky2025critically}.\\

We now proceed to examine the time-evolution of the system under the Hamiltonian:

\begin{equation}
H = \frac{J}{2N} \sum_{i,j=1}^N \sigma_i^z \sigma_j^z + \sum_{n \in \mathbb{Z}} \delta (t-n)\left[\left(\frac{\pi}{m} (1-\epsilon \tau_n)\right) \sum_i \sigma_i^x\right],
\label{eq:TMmodel}
\end{equation}
where $\tau_n$ is the $n$-th element of the quasi-periodic TM sequence. We note that this system would exhibit eternal period-$m$ oscillations when $\epsilon = 0$ and the drive is perfectly periodic for $m=2,3,4$ and $6$. In the remainder of this section, we examine the dynamics of this system when $\epsilon \ne 0$.\\

Before proceeding further, we review salient features of the TM sequence. This sequence is composed of an infinite number of successive subsequences, where the $\mu$-th subsequence, comprises $2^\mu$ elements as follows \cite{thue1906uber,morse1921recurrent}:
\begin{equation}
s (\mu) = s (\mu-1) s^{\rm R}_{1/2} (\mu-1) s^{\rm L}_{1/2} (\mu-1)
\label{tms}
\end{equation}
where, $s^{\rm R}_{1/2} (s^{\rm L}_{1/2})$ denote the right (left) half of the sub-sequence and $s_1 = \{0,1\}$. the first few subsequences of the TMS: 
\begin{eqnarray}
\mu = 1 &:& 0,1 \nonumber\\
\mu = 2 &:& 0,1,1,0 \nonumber \\
\mu = 3 &:& 0,1,1,0,1,0,0,1 \nonumber \\
\mu = 4 &:& 0,1,1,0,1,0,0,1,1,0,0,1,0,1,1,0 \nonumber
\end{eqnarray}
Alternatively, this sequence can be generated by using the substitution rule: $0 \rightarrow 0 1$ and $1 \rightarrow 1 0$; an illustration of this procedure is shown in Fig.~\ref{fig1}(b). The self-similarity of the TM sequence can be seen by observing that removing every second element of the sequence results in the same sequence. This quasi-periodicity of the TM sequence originates from this self-similarity. The aperiodic nature of the drive is evident from the Fourier transform of the drive protocol, $\sum_{n \in \mathbb{Z}} \delta (t-n)\left(\frac{\pi}{m} \epsilon \tau_n\right)$ (for $m=2$ and $\epsilon = 0.01$ and $0.03$) is shown in Fig.~\ref{fig1}(b); the spectrum for other values of $m$ and $\epsilon$ would be qualitatively similar (with a suitably re-scaled y-axis). The Fourier spectrum of the entire drive protocol, $\left(\frac{\pi}{m} (1-\epsilon \tau_n)\right)$ includes an additional Fourier component at $\nu = 0$. \\

We now proceed to characterize the dynamical phases of this model by computing the time-evolution of the z-magnetization. Remarkably, we find that this system can exhibit periodic temporal order even when $\epsilon \ne 0$. We dub this phase a `quasi-DTC', since the underlying drive protocol does not have a discrete time translation symmetry. Thus, although this system behaves almost like a DTC, it can not be called a true DTC, since there is an absence of discrete TTSB. Our nomenclature of `quasi-DTC' is inspired from the term `quasi-many-body localization' that has been used to describe many-body localization-like non-ergodic dynamics in disorder-free systems~\cite{yao2016quasi,barbiero2015out}. \\

\begin{figure*}
		\includegraphics[scale=0.25]{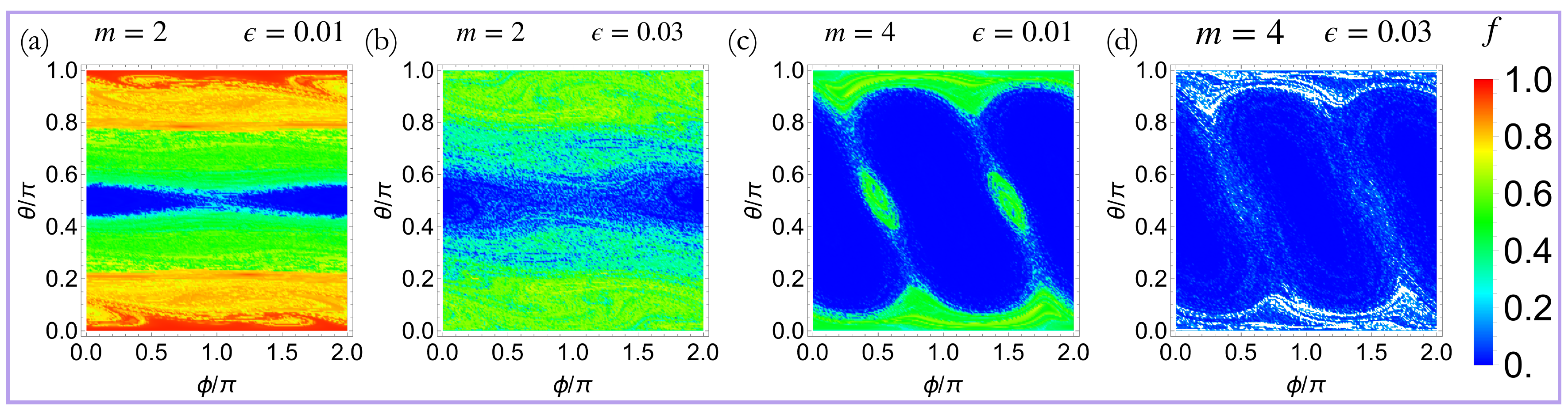}
	\caption{{\bf Crystalline fraction for other initial states:} Density plots of the crystalline fraction, $f$ for initial states of the form $\{S^x (t=0),S^y (t=0),S^z (t=0)\} = \{\sin(\theta)\cos(\phi), \sin(\theta)\sin(\phi), \cos(\theta)\}$. Quasi-DTC behavior is seen for a large class of initial states both for the $m=2$ and $m=4$ quasi-DTCs, with the largest crystalline fraction for the z-polarized initial state. The quasi-DTC fraction, $f$ is larger for $m=2$, compared to $m=4$, thereby signifying that the $m=2$ quasi-DTCs are more robust.}
		\label{fig3}
\end{figure*}

Following previous studies on DTCs~\cite{choi2017observation,he2025experimental}, we establish the robustness of this phase by computing the quasi-DTC fraction:
\begin{equation}
    f= \frac{\sum_{\nu_0} |S(\nu_0)|}{\sum_ {\nu} |S(\nu)|},
    \label{eq:DTCFrac}
\end{equation}
where $S(\nu) = \sum_n S^z(n) \exp(i 2 \pi \nu n)$ is the Fourier transform of $S^{z}$, and $\nu_0$ represent the frequencies characterizing the quasi-DTCs. A quasi-DTC is characterized by sharp peaks at $\nu_0=1/2$ for period-doubling DTCs, and $\nu_0=1/m$ and $\nu_0=1-1/m$ for HO-DTCs (see Fig.~\ref{fig1}(c)). Thus the crystalline fraction $f$ quantifies the normalized weight of these sharp Fourier peaks at $\nu_0$, such that a high crystalline fraction ($f\sim 1$) indicates a robust quasi-DTC, while a low value of $f$ ($f \sim 0$) indicates that the quasi-DTC order has melted. Our calculations reveal that this system can exhibit quasi-DTC order for a wide parameter regime, when $m=2$ and $4$ (see Fig.~\ref{fig2}(a1)-(a4)). Quasi-DTC phases are also seen for $m=3$ and $6$, though these persist over a smaller region of parameter space.\\

Having established the existence of the quasi-DTC phase, we further characterize the dynamical phase diagram of the system by studying the `decorrelator' which is a semiclassical analog of the out-of-time-ordered-correlator~\cite{pizzi2021higher}. The decorrelator measures how the distance between two initially close copies of the system (represented by $\psi$ and $\psi^{\prime}$) increases with time:
\begin{equation}
    d^2 (t) = [\vert \psi_{\uparrow}(t)\vert^2 - \vert \psi^{\prime}_{\uparrow}(t)\vert^2]^2 + [\vert \psi_{\downarrow}(t)\vert^2 - \vert \psi^{\prime}_{\downarrow}(t)\vert^2]^2,
    \label{eq:decorr}
\end{equation}
where we have set
\begin{equation}
\psi_{\uparrow} (0) = 1, \psi_{\downarrow} (0) = 0
\end{equation}
and 
\begin{equation}
\psi^{\prime}_{\uparrow} (0) = \cos(\Delta)e^{-i \phi/2}, \psi^{\prime}_{\downarrow} (0) = \sin(\Delta)e^{-i \phi/2}
\end{equation}
where $\Delta = \phi = 10^{-6}$. The decorrelator captures sensitive dependence on initial conditions, such that a high value of $d^2$ indicates chaos. This kind of chaotic dynamics leads to infinite-temperature thermalization, where the $z-$magnetization, $S^z(t)$ shows small chaotic oscillations with an amplitude $\ll 1$ around 0. On the other hand, a low value of $d^2$ ($d^2 \sim 0$ indicates non-chaotic dynamics corresponding to either a quasi-DTC or a trivial oscillatory phase. Our results shown in Fig.~\ref{fig2}(b1)-(b4) demonstrate that the quasi-DTC phase is characterized by a low value of the time-averaged decorrelator with $\langle d^2 \rangle \sim 0$. Before concluding this discussion, we note that a time-dependent Bogoliubov theory can alternatively be employed to classify non-chaotic and chaotic phases in our model~\cite{bukov2015prethermal,lellouch2017parametric,lemm2025out}.\\

Up to this point, we have only presented he results for the time-evolution of the $z-$polarized initial state. However, in order to qualify as a prethermal phase of matter, the quasi-DTC should exhibit persistent oscillations for a measurable fraction of initial states in the thermodynamic limit. In order to examine this, we study the dynamics of the kicked LMG chain for initial conditions parameterized by $\{S^x (t=0),S^y (t=0),S^z (t=0)\} = \{\sin(\theta)\cos(\phi), \sin(\theta)\sin(\phi), \cos(\theta)\}$. Figure~\ref{fig3} shows that this system exhibits quasi-DTC dynamics for a large class of initial states when $\epsilon \ne 0$. The quasi-DTC fraction, $f$ is the maximum for the $z-$polarized initial state and it is larger for the $m=2$ quasi-DTC. We conclude that the quasi-DTC order is more robust for $m=2$ compared to $m=4$.\\ 

\begin{figure*}
		\includegraphics[scale=0.21]{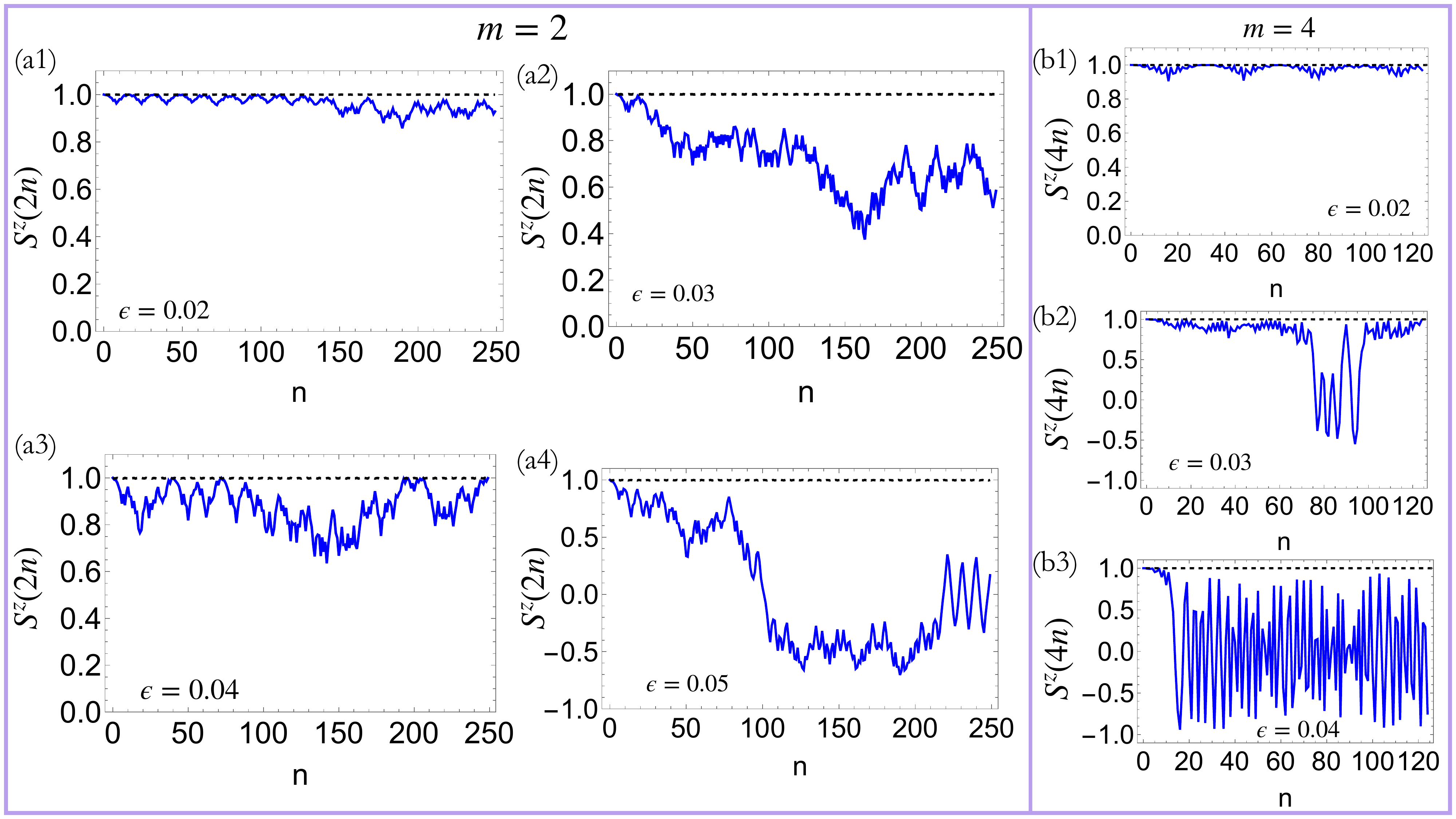}
	\caption{{\bf Comparison of the dynamics of the periodically and Thue-Morse driven LMG chain:} The solid blue and dashed black lines show the time-evolution of the system under a TM drive and the effective Floquet propagator, $\overline{U}$ (eq~\ref{eq:UFloq}) respectively for the period-2 quasi-DTC ((a1)-(a4)) and the period-4 quasi-DTC ((b1)-(b3)). In the small $\epsilon$-regime, the effective Floquet description faithfully captures the dynamics of the TM driven LMG chain. For larger values of $\epsilon$ however, the dynamics of the system deviates significantly from the effective Flqouet description.}
		\label{fig4}
\end{figure*}

We now proceed to elucidate the origin of the prethermal quasi-DTC phase in the small $\epsilon$ regime. We begin by noting that the time evolution operator is composed of a quasiperiodic sequence of two types of effective ``dipoles", $U_{+} = U_0 U_1$ and $U_{-} = U_1 U_0$~\cite{zhao2021random}, where:
\begin{equation}
U_0 =\overline{U} \exp(-i \frac{\pi}{m} \frac{\epsilon}{2} \sum_j \sigma_j^x), U_1=\overline{U} \exp(i \frac{\pi}{m} \frac{\epsilon}{2} \sum_j \sigma_j^x), 
\nonumber
\end{equation}
and
\begin{equation}
\overline{U} = \exp(-i \frac{J}{2N} \sum_{j,l} \sigma_j^z \sigma_l^z) \exp(- i \frac{\pi}{m}(1 -\frac{\epsilon}{2}) \sum_j \sigma_j^x).
\label{eq:UFloq}
\end{equation} 
In the small $\epsilon$ $(\pi \epsilon \ll 1)$ limit, $U_{+} \approx U_{-} \approx \overline{U}^2$, and the evolution of the spin chain is approximately described by the Floquet propagator $\overline{U}$. This simplification leads to a remarkable conclusion: this system can be a quasi-DTC and exhibit long-lived periodic oscillations in the same parameter regimes, where the Floquet LMG chain exhibited DTC dynamics. We have examined the validity of our analysis by comparing the exact dynamics of the model with the effective dynamics generated by $\overline{U}$. As shown in Fig.~\ref{fig4}, in the small $\epsilon$-regime, $\overline{U}$ can provide an effective description of the quasi-DTC dynamics. However, when $\epsilon$ is large, the mapping to the Floquet LMG chain breaks down, and the system thermalizes. \\

\begin{figure*}
		\includegraphics[scale=0.19]{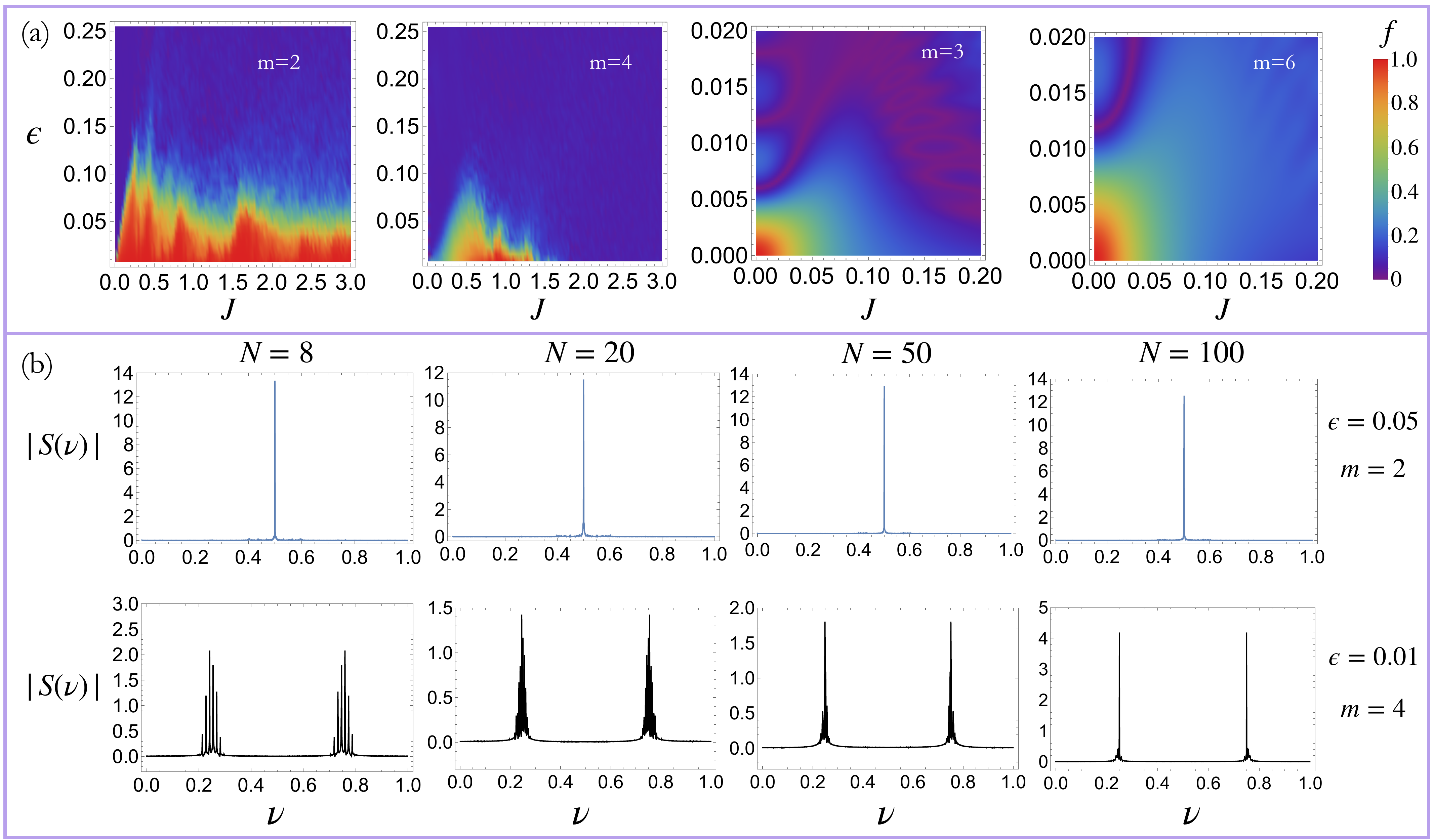}
	\caption{{\bf Exact quantum dynamics of the quasi-DTCs:} (a) The quasi-DTC fraction for an LMG chain composed of 100 spins after 500 periods. Analogous to the thermodynamic limit results shown in Fig.~\ref{fig2}, the period-2 and period-4 quasi-DTCs are exhibit robust periodic temporal order for a wide parameter regime, while the period-3 and period-6 quasi-DTCs exhibit periodic temporal order over a much smaller region of parameter space. (b) The Fourier transform of the z-magnetization, $|S(\nu)|$ for different values of the total number of spins, $N$. When $N$ is small ($\sim 10$), only the period-2 quasi-DTC survives. The period-4 quasi-DTC emerges for larger values of $N$. These calculations have been performed for $J = 0.35$.}
		\label{fig5}
\end{figure*}

Finally, we investigate the effect of quantum fluctuations on quasi-DTCs by studying the dynamics of the system when the number of spins, $N$ is finite. In this case, the total spin operator $S^2 = \overrightarrow{S}\cdot \overrightarrow{S}$ (where $\overrightarrow{S} = \sum_j (\hat{x} \sigma_j^x+\hat{y} \sigma_j^y+\hat{z} \sigma_j^z)$) commutes with $H$, and the dimension of the effective Hilbert space is $N+1$ (instead of $2^N$), thus enabling us to study the exact dynamics for large system sizes.  As shown in Fig.~\ref{fig5}(a), we find that the period-2 and period-4 quasi-DTCs remain stable over a wide parameter regime in the presence of these quantum fluctuations when $N=100$. Furthermore, similar to the $N \rightarrow \infty$ results, the period-3 and period-6 quasi-DTCs are more fragile and they exist within a much smaller region of parameter space. Finally, we note that for much stronger quantum fluctuations ($N \sim 10$), only the period-doubling DTC survives (Fig.~\ref{fig5}(b)); a similar feature is observed for the periodically driven model~\cite{pizzi2021higher}.\\

\section{Potential Experimental Realizations}
\label{sec:Experiment}

The `all-to-all' interacting LMG model can be realized in a variety of quantum emulator platforms, including atom-photon coupled systems~\cite{mivehvar2021cavity,li2022collective,muniz2020exploring}, trapped ion crystals~\cite{jurcevic2017direct,lanyon2011universal}, superconducting qubits~\cite{xu2020probing}, and ultracold bosons in a double-well potential~\cite{shin2004atom,albiez2005direct,carr2010dynamical}. In this section, we discuss possible routes to realize our model (Eq.~\ref{eq:model}) in the context of cavity quantum electrodynamics and trapped ion quantum processors. \\

For the first realization, we study the dynamics of an ensemble of ultracold atoms coupled to a single-mode cavity~\cite{norcia2018cavity,muniz2020exploring}. This system provides a natural route to realize `all-to-all' interactions. The dynamics of this system can be described by a master equation for the density operator $\hat{\rho}$ of the atom-light system:
\begin{equation}
    \frac{d\hat{\rho}}{dt} = -i\left[ \hat{H}_{\mathrm{tot}}, \hat{\rho} \right] + \mathcal{L}_c[\hat{\rho}] \label{eqn:Mastereqn}
\end{equation}
Here, the Hamiltonian $\hat{H}_{\mathrm{tot}}$ is given by: 
\begin{equation}
    \hat{H}_{\mathrm{tot}}  =  \frac{\omega_a}{2} S^z_i + \omega_c\hat{a}^{\dagger}\hat{a} + \Omega_p\left( \hat{a}e^{i\omega_p t} + \hat{a}^{\dagger}e^{-i\omega_p t} \right) +  g  \left( \hat{a}S^+_i + \hat{a}^{\dagger}S^-_i\right) ,
\end{equation}
where $\hat{a}$ represents the annihilation operator of the cavity photons, $\omega_a$ is the atomic transition frequency, $\omega_c$ the cavity mode frequency, $\Omega_p$ ($\omega_p$) the amplitude (frequency) of the driving laser field, $S^+ = S^x+i S^y$, and $S^- = S^x-i S^y$. We note that we will analyze this system in the bad cavity regime where the photon leakage is high, and spontaneous emission can be neglected. \\

The Lindblad term describes the photon leakage from the cavity at rate $\kappa$: 
\begin{equation}
    \mathcal{L}_c[\hat{\rho}] = \kappa \left(2\hat{a}\hat{\rho}\hat{a}^{\dagger} - \hat{a}^{\dagger}\hat{a}\hat{\rho} - \hat{\rho}\hat{a}^{\dagger}\hat{a} \right),
\end{equation}
We now proceed to adiabatically eliminate the photon modes and obtain an effective description of the atomic dynamics. This is obtained by first moving to the rotating frame ( $\hat{a} \to \hat{a}e^{i\omega_p t}$ and $S^+ \to S+ e^{-i\omega_p t}$), such that we obtain
\begin{equation}
    \frac{d\hat{\rho}}{dt} = -i\left[ \hat{H}_{\mathrm{eff}}, \hat{\rho} \right] + \mathcal{L}_c[\hat{\rho}],
\end{equation}
where 

\begin{equation}
    \hat{H}_{\mathrm{eff}} = (\Delta - \delta)\hat{a}^{\dagger}\hat{a} + (\Omega_p^*\hat{a} + \Omega_p\hat{a}^{\dagger}) -\frac{\delta}{2} S^z +  g \left( \hat{a}\hat{S}^+ + \hat{a}^{\dagger}\hat{S}^-_i \right).
\end{equation}

Here $\Delta=\omega_c-\omega_a$ is the cavity detuning and $\delta=\omega_p-\omega_a$ is the atomic detuning. Next, we can write down the equations of motion for both the collective spin operators, $S^{\alpha}$ and the photon annihilation operator, $a$, when the atomic detuning. In the adiabatic approximation, we set $\partial_t a = 0$, resulting in the following equation for the photonic field operator:
\begin{equation}
    a(t) = \frac{g}{(\Delta-\delta) - i \kappa} S^{-}.
\end{equation}
We are next going to work in the limit of small atomic detuning, $\delta \approx 0 $. This results in an effective spin Hamiltonian, $\hat{H}$ that governs the time-evolution of the reduced density matrix of the atoms $\hat{\rho}_s$:
\begin{equation}
    \frac{d\hat{\rho}_s}{dt} = -i\left[ \hat{H}, \hat{\rho}_a \right], 
\end{equation}
with 
\begin{equation}
    \hat{H} = \chi S^+S^- + \frac{\Omega}{2} S^x  , 
\end{equation}
for $\chi = - g^2 \Delta /[\Delta^2 + \kappa^2]$ and $\Omega = -2g\vert\Omega_p\vert \Delta/[\Delta^2 + \kappa^2$. \\

These expressions simplify considerably when $\kappa \ll \vert \Delta \vert$, such that $\chi \approx -g^2 /\Delta$ and $\Omega = -2g\vert\Omega_p\vert\Delta$. In the fully symmetric subspace, this Hamiltonian takes an even simpler form:
\begin{equation}
    \hat{H} = \chi \frac{N}{2}\bigg(\frac{N}{2}+1 \bigg) - \chi (S^z)^2 + \Omega S^x, 
\end{equation}
Thus, for a fixed $N$ apart from a constant offset, $\hat{H}$ has the same form as our model in Eq.~\ref{eq:model} with $\chi = - 2J/N$. We conclude that our model can be realized by periodically modulating the pump laser field amplitude, $\Omega_p$. \\

An alternative route to realize this model is to employ trapped-ion quantum emulators, where pseudo-spin-1/2 ions are confined in a one-dimensional Paul trap. Long-range interactions are engineered in this system via collective transverse vibrations of the chain that couple to the ions via Raman lasers~\cite{blatt2012quantum,monroe2021programmable}. The Hamiltonian describing this system is:
\begin{equation}
    H_1 = \sum_{i,j} \frac{J}{\vert j-i \vert^{\alpha}} \sigma^{x}_i \sigma^{x}_j,
\end{equation}
where $0 \le \alpha \le 3$. In particular, Jurcevic {\it et al.} have realized the $\alpha \approx 0$ regime of this model with a one-dimensional crystal of ${\rm ^{40}Ca^{+}}$ ions~\cite{jurcevic2017direct}. Furthermore, arbitrary time-dependent effective magnetic fields are regularly engineered in these experiments by controlling the lasers that couple the collective vibrational modes to the ions~\cite{monroe2017nature,kyprianidis2021observation}. Thus, our predictions can be verified in state-of-the-art quantum simulator platforms.\\

\section{Summary and Outlook}
\label{sec:Summary}
Time crystals epitomize the profound principle of TTSB in many-body systems. In this work, we have demonstrated that periodic temporal order akin to time crystals can emerge in quasi-periodically driven systems. This new phase - dubbed a quasi-DTC - is characterized by the spontaneous breaking of the average discrete time-translation symmetry. We establish the existence of both period-doubling and `higher-order' quasi-DTCs and demonstrate that they originate from the interplay of the Thue-Morse driving and long-range interactions. We emphasize that quasi-DTCs are qualitatively distinct from MBL-DTCs, whose robustness is guaranteed only when the drive is perfectly periodic \cite{sondhi2016prb}. Although most of our calculations have been in the thermodynamic limit, we have demonstrated that these quasi-DTCs persist even in the presence of quantum fluctuations in a finite-size chain. Finally, we have discussed possible experimental realizations of this model. Our work clearly demonstrates that quasiperiodically driven many-body systems can host a rich array of non-equilibrium phases.\\

There are several possible extensions of this work. One promising direction is to examine routes to realize a quasi-DTC in the presence of dissipation \cite{kessler2020observation}. This can be an intriguing direction of research, since strong dissipation can prevent heating~\cite{cosme2023bridging}, but it also inevitably introduces noise that can destroy the quasi-DTC order~\cite{zhu2019dicke}. More generally, it would be interesting to explore the dynamics of long-range interacting spin chains subjected to different forms of quasiperiodic drive \cite{ray2019dynamics,pilatowsky2023complete}, as well as dissipation \cite{schutz2016dissipation}. Alternatively, it would be intriguing to investigate schemes to leverage quasi-DTCs for quantum-enhanced metrology~\cite{yousefjani2025discrete,biswas2025floquet} and other quantum information processing protocols~\cite{carollo2020nonequilibrium,estarellas2020simulating,engelhardt2024unified}. Finally, studying the fate of DTCs in the presence of completely noisy driving as well as other kinds of aperiodic driving protocols would be an interesting direction for future research.\\

\section*{Acknowledgements}
The authors want to thank Biao Huang for helpful discussions. This work is supported by the AFOSR Grant No. FA9550-23-1-0598, the MURI-ARO Grant No. W911NF17-1-0323 through UC Santa Barbara, the Shanghai Municipal Science and Technology Major Project through the Shanghai Research Center for Quantum Sciences (Grant No. 2019SHZDZX01), and the DST SERB project SRG/2023/002730.

\bibliographystyle{apsrev4-2}
\bibliography{ref} 

\end{document}